\documentclass[prd,showpacs,amsmath,amssymb]{revtex4}

\usepackage{graphicx}
\usepackage{dcolumn}
\usepackage{bm}
\usepackage{psfrag}
\usepackage{subfigure}

\newcommand{\be}{\begin{eqnarray}}
\newcommand{\ee}{\end{eqnarray}}
\newcommand{\bn}{\begin{eqnarray*}}
\newcommand{\en}{\end{eqnarray*}}

\newcommand{\nn}{\nonumber \\}
\newcommand{\nl}{\\}

\renewcommand{\vec}[1]{\mbox{\boldmath$#1$}}
\renewcommand{\d}{\mbox{\rm d}}

\renewcommand{\th}{\ensuremath{\theta}}
\newcommand{\vth}{\ensuremath{\vartheta}}
\newcommand{\Th}{\ensuremath{\Theta}}
\newcommand{\ph}{\ensuremath{\phi}}
\newcommand{\Ph}{\ensuremath{\Phi}}
\newcommand{\vph}{\ensuremath{\varphi}}
\newcommand{\al}{\ensuremath{\alpha}}
\newcommand{\bt}{\ensuremath{\beta}}
\newcommand{\sg}{\ensuremath{\sigma}}
\newcommand{\gm}{\ensuremath{\gamma}}
\newcommand{\dl}{\ensuremath{\delta}}
\newcommand{\lm}{\ensuremath{\lambda}}
\newcommand{\Lm}{\ensuremath{\Lambda}}

\newcommand{\Dl}{\ensuremath{\Delta}}
\newcommand{\Sg}{\ensuremath{\Sigma}}
\newcommand{\Gm}{\ensuremath{\Gamma}}

\newcommand{\omK}{\ensuremath{\omega_{\rm K}}}

\newcommand{\Xvec}{\ensuremath{\vec{X}}}

\newcommand{\nabvec}{\ensuremath{\vec{\nabla}}}

\newcommand{\lt}{\ensuremath{\left}}
\newcommand{\rt}{\ensuremath{\right}}

\renewcommand{\d}{\mbox{\rm d}}

\begin{document}

\pagenumbering{arabic}

\title{Tidal Dynamics in Cosmological Spacetimes}%

\author{Bahram Mashhoon}
\email{mashhoonb@missouri.edu}
\affiliation{%
Department of Physics and Astronomy, University of Missouri-Columbia \\
Columbia, Missouri 65211, USA
}%
\author{Nader Mobed}
\email{nader.mobed@uregina.ca}
\author{Dinesh Singh}
\email{singhd@uregina.ca}
\affiliation{%
Department of Physics, University of Regina \\
Regina, Saskatchewan, S4S 0A2, Canada
}%
\date{\today}

\begin{abstract}

We study the relative motion of nearby free test particles in cosmological spacetimes,
such as the FLRW and LTB models.
In particular, the influence of spatial inhomogeneities on local tidal accelerations is investigated.
The implications of our results for the dynamics of the solar system are briefly discussed.
That is, on the basis of the models studied in this paper, we estimate the tidal influence of the cosmic gravitational field
on the orbit of the Earth around the Sun and show that the corresponding temporal rate of
variation of the astronomical unit is negligibly small.

\end{abstract}


\maketitle

\section{Introduction}
\label{Sec1}

The gravitational influence of distant galaxies extends over all of the bodies in our cosmic neighbourhood;
therefore, the relative motion of these bodies would be affected by the tidal field of distant masses.
In general relativity, it is natural to study the relative motion of a test particle with respect to an observer
in the quasi-inertial Fermi co-ordinate system \cite{Synge} established about the worldline of the observer.
To simplify matters, let us assume that the observer follows a geodesic worldline ${\cal C}$ and let $\lm^\mu{}_{(\al)}$
be the observer's local orthonormal tetrad frame that is parallel transported along ${\cal C}$.
Here $\lm^\mu{}_{(0)} = \d x^\mu/\d \tau$ is the observer's local temporal axis and $\tau$ is its proper time,
while $\lm^\mu{}_{(i)}$, $i = 1, 2, 3,$ are unit gyro directions that constitute the observer's local spatial frame.
We choose units such that $c=1$ throughout this paper.
An event $P$ in the neighbourhood of the observer has Fermi co-ordinates $X^\mu = \lt(T, \vec{X}\rt)$;
in fact, $P$ can be orthogonally connected to ${\cal C}$ at $P_0$ via a unique spacelike geodesic of proper length $\sg$
such that $\sg = 0$ at $P_0$.
Let the proper time along ${\cal C}$ at $P_0$ be $\tau$ and $\xi^\mu = \lt(\d x^\mu/\d \sg\rt)_0$ be the unit tangent vector
to the spacelike geodesic at $P_0$; then the Fermi co-ordinates of the event $P$ are defined by
$T = \tau$ and $X^i = \sg \, \xi^\mu \, \lm_\mu{}^{(i)}$.
It follows that the observer permanently occupies the spatial origin of this Fermi co-ordinate system.

The spacetime metric in Fermi co-ordinates can be expressed as
\be
{}^F{}g_{00} & = & -1 - {}^F{}R_{0i0j}(T) \, X^i \, X^j + \cdots,
\label{F-g00}
\nl
{}^F{}g_{0i} & = & -{2 \over 3} {}^F{}R_{0jik}(T) \, X^j \, X^k + \cdots,
\label{F-g0i}
\nl
{}^F{}g_{ij} & = & \dl_{ij} - {1 \over 3} {}^F{}R_{ikjl}(T) \, X^k \, X^l + \cdots,
\label{F-gij}
\ee
where
\be
{}^F{}R_{\al \bt \gm \dl}(T) & = & R_{\mu \nu \rho \sg} \, \lm^\mu{}_{(\al)} \, \lm^\nu{}_{(\bt)} \, \lm^\rho{}_{(\gm)} \, \lm^\sg{}_{(\dl)}
\label{F-Riemann}
\ee
is the Riemann curvature tensor along ${\cal C}$ projected on the tetrad frame of the observer.
The Fermi co-ordinates are admissible in a cylindrical spacetime region around ${\cal C}$ with $|\vec{X}| < {\cal L}$,
where ${\cal L}$ is a measure of the radius of curvature of spacetime.

The motion of a free test particle is given by
\be
{\d^2 X^\mu \over \d s^2} + {}^F{}\Gm_{\al \bt}^\mu \, {\d X^\al \over \d s} \, {\d X^\bt \over \d s} & = & 0,
\label{geodesic-eq}
\ee
where $\d X^\mu / \d s = \Gm \lt(1, \vec{V}\rt)$ is the particle's timelike four-velocity vector.
Equation (\ref{geodesic-eq}) can be expressed as
\be
{1 \over \Gm} \, {\d \Gm \over \d T} & = & - {}^F{}\Gm_{\al \bt}^0 \, {\d X^\al \over \d T} \, {\d X^\bt \over \d T}
\label{geodesic-eq-1}
\ee
and the reduced geodesic equation
\be
{\d^2 X^i \over \d T^2} + \lt({}^F{}\Gm_{\al \bt}^i - {}^F{}\Gm_{\al \bt}^0\, V^i \rt){\d X^\al \over \d T} \, {\d X^\bt \over \d T} & = & 0.
\label{geodesic-eq-2}
\ee
This latter equation, to linear order in distance away from $\cal C$, is the {\it generalized Jacobi equation} \cite{Chicone1}
\be
\lefteqn{{\d^2 X^i \over \d T^2} + {}^F{}R_{0i0j} \, X^j + 2 \, {}^F{}R_{ikj0} \, V^k \, X^j }
\nn
&& + {2 \over 3} \lt(3 \, {}^F{}R_{0kj0} \, V^i \, V^k + {}^F{}R_{ikjl} \, V^k \, V^l + {}^F{}R_{0kjl} \, V^i \, V^k \, V^l\rt) X^j \ = \ 0,
\label{generalized-Jacobi}
\ee
and the modified Lorentz factor is given by
\be
{1 \over \Gm^2} & = & 1 - V^2 + {}^F{}R_{0i0j} \, X^i \, X^j + {4 \over 3} \, {}^F{}R_{0jik} \, X^j \, V^i \, X^k
\nn
& &{} + {1 \over 3} \, {}^F{}R_{ikjl} \, V^i \, X^k \, V^j \, X^l.
\label{Lorentz-factor}
\ee
Equation (\ref{generalized-Jacobi}) reduces to the Jacobi equation when the velocity-dependent
terms are negligibly small.

In some situations of physical interest, it is possible to have purely one-dimensional motion, say along the $Z$-direction.
In this case, equation (\ref{generalized-Jacobi}) reduces to
\be
{\d^2 Z \over \d T^2} + k(T) (1 - 2 \dot{Z}^2) Z & = & 0,
\label{generalized-Jacobi-1D}
\ee
where $\dot{Z} = \d Z / \d T$ and $k(T) = {}^F{}R_{TZTZ}$.
For $|\dot{Z}| \ll 1$, equation (\ref{generalized-Jacobi-1D}) reduces to the standard Jacobi equation.
However, when the relative speed cannot be neglected in comparison with the speed of light, the generalized Jacobi equation (\ref{generalized-Jacobi-1D})
has solutions with $\dot{Z} = \pm 1/\sqrt{2}$ such that the relative motion within the linear approximation scheme is uniform at the {\it critical speed}
$V_c = 1/\sqrt{2}$.
Below this speed, the relative motion should essentially conform to expectations based on the post-Newtonian approximation.
On the other hand, novel relativistic tidal effects are expected to occur above the critical speed.

Let the critical solutions of equation (\ref{generalized-Jacobi-1D}) be expressed as $Z_c(T) = Z_i \pm V_c (T - T_i)$, where
$T_i$ is an initial time and $Z_i = Z_c(T_i)$.
Then, the behaviour of $Z(T)$ near a critical solution can be examined by introducing $\zeta(T) = Z - Z_c$.
It follows from (\ref{generalized-Jacobi-1D}) that
\be
{\d^2 \zeta \over \d T^2} - 2 \, k(T) \lt(\dot{\zeta}^2 \pm 2 V_c \, \dot{\zeta}\rt) \lt(\zeta + Z_c\rt) & = & 0.
\label{W0}
\ee
Keeping only terms of linear order in (\ref{W0}), we find
\be
{1 \over \dot{\zeta}} \, {\d \dot{\zeta} \over \d T} & = & \pm 4 \, V_c \, Z_c \, k(T),
\label{W1}
\ee
where $\dot{\zeta} = \d \zeta / \d T$.
It follows that the long-term behaviour of this motion can be determined from
\be
{\cal I} & = & \int_{T_i}^\infty \lt(T + C_0\rt) k(T) \, \d T,
\label{I}
\ee
where $C_0 = -T_i \pm \sqrt{2} \, Z_i$ and $\pm \, 2 \, V_c \, Z_c = T + C_0$.
For instance, the critical solutions are attractors for ${\cal I} = -\infty$ \cite{Chicone1}.
Equation (\ref{generalized-Jacobi-1D}) has been extensively studied in black hole spacetimes in connection with
the problem of astrophysical jets \cite{Chicone2,Chicone3}.

The main purpose of this paper is to study equation (\ref{generalized-Jacobi-1D}) for some cosmological models.
It turns out that for the standard FLRW models, $k(T)$ is characterized by the {\it deceleration} of the universe.
This point is demonstrated in the following section.
In section~\ref{Sec3}, a locally inhomogeneous model is employed to determine the corresponding $k(T)$.
In this way, the influence of cosmic spatial inhomogeneities on local measurements is elucidated.
This treatment is extended in section~\ref{Sec4} to the Lema\^{i}tre-Tolman-Bondi (LTB) models.
The consequences of the results of section~\ref{Sec4} for the dynamics of the solar system are briefly discussed in
section~\ref{Sec5} and Appendix~\ref{Appendix:A}.
Section~\ref{Sec6} contains a discussion of our results.

\section{FLRW Models}
\label{Sec2}

Let us first consider the FLRW spacetimes given in isotropic co-ordinates $x^\mu = \lt(t, x^i\rt)$ by
\be
-\d s^2 & = & -\d t^2 + {S^2(t) \over f^2(r)} \, \dl_{ij} \, \d x^i \, \d x^j \, ,
\label{FLRW-line-element}
\ee
where $S(t) > 0$ is the scale factor and
\be
f(r) & = & 1 + {1 \over 4} \, \kappa \, r^2
\label{f}
\ee
with $\kappa = -1, 0$, or $+1$ for the open, flat, or closed universe models, respectively.
We note that the spatial co-ordinates $x^i$ are dimensionless, while the scale factor $S$ has the dimension of length.

A fundamental observer in this spacetime occupies a fixed position in space and follows a geodesic worldline $\cal C$
with proper time $\tau = t$.
Let us choose a fundamental observer with $\lm^\mu{}_{(0)} = \dl^\mu{}_0$ and $\lm^\mu{}_{(i)} = \lt[f(r)/S(t)\rt]\dl^\mu{}_i$,
which can be shown to be parallel transported along the geodesic worldline.
The curvature tensor ${}^F{}R_{\al \bt \gm \dl}$ along $\cal C$ can be expressed as a $6 \times 6$ matrix $\lt({}^F{}{\cal R}_{JK}\rt)$,
where the indices $J$ and $K$ range over the set \{01, 02, 03, 23, 31, 12\}.
Then,
\be
{{}^F{}\cal R} \ = \
\left(%
\begin{array}{cc}
  E & B \\
  B^\dag & N \\
\end{array}%
\right),
\label{R-matrix}
\ee
where for the FLRW models $E = u(t) \, I$, $B = 0$, and $N = w(t) \, I$.
Here $I$ is the $3 \times 3$ identity matrix and
\be
u(t) & = & -{1 \over S} \, {\d^2 S \over \d t^2} \, ,
\label{u-def}
\nl
w(t) & = & {1 \over S^2} \lt(\d S \over \d t\rt)^2 + {\kappa \over S^2} \, .
\label{w-def}
\ee
We note in passing that (\ref{u-def}) and (\ref{w-def}) imply
\be
S \, {\d w \over \d t} & = & - 2 \lt(u + w\rt) {\d S \over \d t} \, .
\label{dw/dt}
\ee

It is possible to show that the only nonzero components of the Einstein tensor, $G_{\mu \nu} = R_{\mu \nu} - {1 \over 2} \, g_{\mu \nu} \, R$,
are given by
\be
G_{\mu \nu} \, \lm^\mu{}_{(0)} \, \lm^\nu{}_{(0)} & = & 3 \, w(t) \, ,
\label{G00-tetrad}
\nl
G_{\mu \nu} \, \lm^\mu{}_{(i)} \, \lm^\nu{}_{(j)} & = & \lt[2 \, u(t) - w(t) \rt] \dl_{ij} \, .
\label{Gij-tetrad}
\ee
The gravitational field equations (with a cosmological constant $\Lm$) are given by
\be
G_{\mu \nu} + \Lm \, g_{\mu \nu} & = & 8 \pi G \, T_{\mu \nu} \, ,
\label{Einstein+Lm}
\ee
where $T_{\mu \nu}$ is the stress-energy tensor for a perfect fluid
\be
T_{\mu \nu} & = & \mu(t) \, u_\mu \, u_\nu + p(t) \lt(g_{\mu \nu} + u_\mu \, u_\nu\rt).
\label{T}
\ee
Here, $\mu(t)$ and $p(t)$ are the invariant density and pressure of the cosmic fluid, and the fundamental observers
are assumed to be comoving with the cosmic fluid, i.e. $u^\mu = \lm^\mu{}_{(0)}$.
Equations (\ref{G00-tetrad})--(\ref{T}) imply that $u$ and $w$, defined respectively by (\ref{u-def}) and (\ref{w-def}),
are given by
\be
u(t) & = & {4 \pi G \over 3} \lt(\mu + 3 \, p\rt) - {\Lm \over 3} \, ,
\label{u+Lm}
\nl
w(t) & = & {8 \pi G \over 3} \, \mu + {\Lm \over 3} \, .
\label{w+Lm}
\ee
It follows from (\ref{dw/dt}), (\ref{u+Lm}), (\ref{w+Lm}) and the constancy of $\Lm$ that
\be
{1 \over 3} \, {\d \mu \over \d t} & = & -{1 \over S} \, {\d S \over \d t} \lt(\mu + p\rt).
\label{dm/dt}
\ee
This is the expression of the first law of thermodynamics for the adiabatic flow of the perfect fluid under consideration here.
The standard Hubble and deceleration parameters ($H$ and $q$) are given by
\be
H & = & {1 \over S} \, {\d S \over \d t} \, , \qquad q \, H^2 \ = \ -{1 \over S} \, {\d^2 S \over \d t^2} \, ,
\label{Hubble-deceleration}
\ee
so that $u = q \, H^2$ and $w = H^2 + \kappa/S^2$.

A Fermi coordinate system $(T, \Xvec)$ can be established along the worldline of a
fundamental observer; then, the observer has Fermi coordinates $(T, \vec{0})$, where $T = t$.
The FLRW spacetime metric in spherical Fermi co-ordinates
\be
X & = & \rho \, \sin \Th \, \cos \Ph \, , \qquad Y \ = \ \rho \, \sin \Th \, \sin \Ph \, , \qquad Z \ = \ \rho \, \cos \Th \, ,
\label{Fermi-coord}
\ee
can be expressed as \cite{Mashhoon1}
\be
- \d s^2 & = & -\lt[1 + u(T) \rho^2\rt] \d T^2 + \d \rho^2 + \lt[1 - {1 \over 3} \, w(T) \, \rho^2 \rt] \rho^2 \lt(\d \Th^2 + \sin^2 \Th \, \d \Ph^2\rt).
\label{FLRW-fermi}
\ee
This is based on (\ref{F-g00})--(\ref{F-gij}) and thus holds only to second order in distance away from the observer.
It appears that explicit Fermi co-ordinates can be constructed for FLRW models based on the results of Ref.~\cite{Chicone3};
however, the scale factor $S(t)$ must then be explicitly specified.  To have a general treatment for arbitrary $S(t)$,
we must use the implicit approach
as in (\ref{F-g00})--(\ref{F-gij}).

The spatial isotropy of the FLRW models implies that an equation of motion of the form of equation (\ref{generalized-Jacobi-1D})
is possible along any direction in space with $k(T) = u(T) = q H^2$.
Thus the nature of such motion is characterized by the sign of the deceleration parameter.
Imagine, for instance, a free test particle moving radially away from the observer.
For an initially infrarelativistic particle with recession speed less than $V_c$, the particle decelerates (accelerates)
for $q > 0$ ($q < 0$), as expected.
However, for an initially ultrarelativistic particle  with recession speed above $V_c$, the particle accelerates (decelerates)
for $q > 0$ ($q < 0$).
In the following sections, we simply concentrate on the nature of $k(T)$ in certain inhomogeneous models.

In the slow-motion approximation, it is possible to extract a ``Newtonian'' gravitational potential \cite{Mashhoon1}
\be
{\cal V}_{\rm N} & = & {1 \over 2} \, q H^2 \rho^2
\label{V-N}
\ee
from (\ref{FLRW-fermi}).
Equation (\ref{V-N}) has been employed in the discussion of the influence of cosmology on local phenomena \cite{Cooperstock}.
Ref.~\cite{Cooperstock} contains only a partial list of papers on this subject; for background material and further references,
see~\cite{Krasinski}.
The form of the quadratic potential (\ref{V-N}) suggests that the influence of the cosmic gravitational field is
reflected in local experiments---in the solar system, for instance---via a relative tidal acceleration of the form
$\vec{g}_{\rm cosmos} = -q H^2 \vec{X}$.
It is interesting to note that the deceleration parameter $q$ also appears in the second-order expansion of luminosity
distance $d_L$ versus redshift $z$; that is
\be
d_L & = & {1 \over H} \, z + {1 \over 2H} \, (1 - q) \, z^2 + \cdots \, .
\label{d-L0}
\ee
This ``degeneracy'' is removed by spatial inhomogeneities as demonstrated in the following section.

It appears highly likely that instead of providing evidence for dark energy, the observational data from type Ia
supernovae have demonstrated the inadequacy of the standard spatially homogeneous FLRW models of the universe.
That is, spatial inhomogeneities must be taken into account when comparing observational data with theoretical
models of the universe.
This is based on the fact that spatial inhomogeneities mimic dark energy \cite{Celerier1}.
A recent useful review of this topic is contained in \cite{Celerier2}.

\section{Inhomogeneous Models}
\label{Sec3}

It is important to investigate how spatial inhomogeneities in the universe are reflected in local spacetime measurements.
We begin with a metric of the form \cite{Mashhoon1,Partovi}
\be
-\d s^2 & = & - a^2(t,r) \, \d t^2 + b^2(t,r) \, \d r^2 + {\cal R}^2(t,r) \lt(\d \th^2 + \sin^2 \th \, \d \ph^2\rt),
\label{FLRW-inhomogeneous}
\ee
where
\be
a(t,r) & = & 1 + {1 \over 2} \, \al(t) \, r^2 + \cdots \, ,
\label{a}
\nl
b(t,r) & = & \lt[1 + {1 \over 2} \, \bt(t) \, r^2 + \cdots \rt] S(t) \, ,
\label{b}
\nl
{\cal R}(t,r) & = & r\lt[1 + {1 \over 2} \, \gm(t) \, r^2 + \cdots \rt] S(t) \, .
\label{R-FLRW}
\ee
Here, $\al$, $\bt$, and $\gm$ are parameters in the expansion of the cosmic gravitational potentials in powers of the
dimensionless radial co-ordinate $r$ away from the centre of spherical symmetry $(r = 0)$.
This model has been discussed at length \cite{Mashhoon1,Partovi}; in particular, it has been shown in \cite{Partovi} that this
local inhomogeneous model is generally valid for a cosmic fluid that has pressure and satisfies a reasonable
equation of state of the form $p = p(\mu)$ with $p \geq 0$, $\mu \geq 0$ and $\mu \geq 3 p$ or $\mu \geq p$.
We note that at the order of approximation indicated in (\ref{a})--(\ref{R-FLRW}), the homogeneous models of the previous
section are recovered for $\al = 0$ and $\bt = \gm = -\kappa/2$.

We imagine, as before, observers that are fixed in space and carry orthonormal tetrad frames along their worldlines.
The natural tetrad of such an observer is diagonal with nonzero elements
$\lm^0{}_{(0)} = a^{-1}$, $\lm^1{}_{(1)} = b^{-1}$, $\lm^2{}_{(2)} = {\cal R}^{-1}$, and $\lm^3{}_{(3)} = \lt({\cal R} \, \sin \th\rt)^{-1}$.
We are interested in a Fermi co-ordinate system along the worldline $\cal C$ of the observer that is at the centre
of spherical symmetry.
It can be shown explicitly using equations (\ref{FLRW-inhomogeneous})--(\ref{R-FLRW}) that $\cal C$ is a geodesic and the
spatial frame is indeed parallel propagated along $\cal C$ in the $r \rightarrow 0$ limit.
Alternatively, one can employ local Cartesian co-ordinates as in the previous section.
In any case, the curvature components ${}^F R_{\al \bt \gm \dl}$ along $\cal C$ can be explicitly calculated and the results
can be expressed as a $6 \times 6$ matrix (\ref{R-matrix}) with
\be
E & = & U(t) \, I \, , \qquad B \ = \ 0 \, , \qquad N \ = \ W(t) \, I \, ,
\label{FLRW-inhomo-components}
\ee
where for this inhomogeneous model
\be
U(t) & = & q H^2 + {\al(t) \over S^2} \, ,
\label{U}
\nl
W(t) & = & H^2 + {\bt(t) - 3 \gm(t) \over S^2} \, .
\label{W}
\ee
Here, $q$ and $H$ are defined in terms of the scale factor $S(t)$ as in (\ref{Hubble-deceleration}).
Moreover, the analogues of equations (\ref{u+Lm}) and (\ref{w+Lm}) hold for $\mu(t,r)$ and $p(t,r)$ evaluated at $r = 0$;
that is,
\be
U(t) & = & {4 \pi G \over 3} \lt[\mu(t,0) + 3 p(t,0)\rt] - {\Lm \over 3} \, ,
\label{u+Lm1}
\nl
W(t) & = & {8 \pi G \over 3} \, \mu(t,0) + {\Lm \over 3} \, .
\label{w+Lm1}
\ee

The spacetime metric in spherical Fermi co-ordinates about $\cal C$ has the same form as equation (\ref{FLRW-fermi}),
except that $u(t) \rightarrow U(t)$ and $w(t) \rightarrow W(t)$; moreover, $t = T$, since $r = 0$ along $\cal C$.
Let us note in this connection that the quantity $\bt - 3 \, \gm$ in (\ref{W}) is related to the spatial curvature
and reduces to $\kappa$ in the homogeneous limit \cite{Mashhoon1,Partovi}.
The motion of a free test particle in the neighbourhood of $\cal C$ can be studied as in the previous section
except that in this case $k(T) = U(T)$ in equation (\ref{generalized-Jacobi-1D}).
Furthermore, the influence of the cosmic gravitational field on local physics in the slow-motion approximation
appears as a ``Newtonian'' acceleration of the form
\be
\vec{g}_{\rm cosmos} & = & - H^2 \lt(q - C\rt) \Xvec \, ,
\label{g-cosmos}
\ee
where $C(t)$ is the {\it inhomogeneity parameter} given by
\be
C & = & -{\al \over \lt(SH\rt)^2} \, .
\label{C}
\ee
It is interesting to note that in this inhomogeneous model, the luminosity distance-redshift relation
can be expressed as \cite{Mashhoon1,Partovi}
\be
d_L & = & {1 \over H} \, z + {1 \over 2 H} \lt[1 - \lt(q + C\rt)\rt]z^2 + \cdots \, ,
\label{d-L}
\ee
so that the measured ``deceleration'' parameter is in fact $q + C$.
It follows that both $q$ and $C$ could be determined from observation if it were possible to measure $k(T) = H^2 (q - C)$ as well.
It follows from (\ref{u+Lm1}) that
\be
H^2(q - C) & = & 4 \pi G \lt[{1 \over 3} \, \mu(t,0) + p(t,0) \rt] - {\Lm \over 3} \, .
\label{H^2(q-c)}
\ee
Hence, in the absence of a cosmological constant we have $C < q$, which is a noteworthy inequality.

Finally, let us note that by imposing the comoving co-ordinate condition \cite{Partovi}
\be
{\d \over \d t} \, \lt(\bt - 3 \gm\rt) & = & -2 H(t) \, \al(t) \, ,
\label{inhomo-condition-time}
\ee
one can show from (\ref{u+Lm1}) and (\ref{w+Lm1}) that
\be
{\partial \over \partial t} \, \mu(t,0) & = & -3 H(t) \lt[\mu(t,0) + p(t,0) \rt],
\label{inhomo-dm/dt}
\ee
which is the analogue of (\ref{dm/dt}) in this inhomogeneous case.

\section{LTB Models}
\label{Sec4}

Imagine a spacetime metric of the form (\ref{FLRW-inhomogeneous}) with
\be
a(t,r) & = & 1 \, , \qquad b(t,r) \ = \ {{\cal R}'(t,r) \over \sqrt{1 + 2{\cal E}(r)}} \, ,
\label{LTB-a,b}
\ee
where ${\cal R}' = \partial {\cal R}/\partial r$.
It turns out that the gravitational field equations (\ref{Einstein+Lm}) are satisfied in this case for pure dust
\be
T_{\mu \nu} & = & \mu(t,r) \, u_\mu \, u_\nu
\label{Tuv-dust}
\ee
in comoving co-ordinates (i.e. $u^\mu = \dl^\mu{}_0$) provided
\be
{\cal E}(r) & = & {1 \over 2} \, \dot{\cal R}^2 - {G M(r) \over {\cal R}} - {\Lm \over 6} \, {\cal R}^2 \, ,
\label{E-def}
\nl
{\d M(r) \over \d r} & = & 4 \pi \, \mu(t,r) \, {\cal R}^2 \, {\cal R}' \, .
\label{dM/dr-def}
\ee
Here, ${\cal E}(r)$ with ${\cal E} > -1/2$ has the interpretation of the net energy per unit mass of the
spherical shell of dust at radius $r$, $M(r)$ has the interpretation of mass within a sphere of radius $r$,
and $\dot{\cal R} = \partial {\cal R} / \partial t$.
These spherically symmetric inhomogeneous dust models were first discovered by Lema\^{i}tre \cite{Lemaitre}
and further studied by Tolman \cite{Tolman} and Bondi \cite{Bondi}; see \cite{Plebanski} for a detailed discussion.

It is a general result that in a spacetime with a metric of the form $-\d t^2 + g_{ij}(t,x^k) \, \d x^i \, \d x^j$,
any test particle that is at rest in space follows a geodesic \cite{Chicone1}.
Therefore, the fundamental comoving observers in the LTB model follow geodesics.
Moreover, each fundamental observer carries the standard orthonormal tetrad frame (discussed in the previous section)
that has its axes along the directions of the co-ordinates employed in (\ref{FLRW-inhomogeneous}).
It can be shown explicitly that such a tetrad frame is parallel transported along the geodesic path of a fundamental observer.
The off-center cosmological measurements of such observers have been discussed in \cite{Humphreys}.
We are interested, however, in the local tidal dynamics of nearby test particles.

It proves convenient, for the sake of simplicity, to orient the spatial axes of the Fermi frame such that the $X$-axis
points along the polar $(\th)$ direction, the $Y$-axis points along the azimuthal $(\ph)$ direction, and the $Z$-axis
points along the radial $(r)$ direction; that is, for the purposes of this section we choose $\lm^\mu{}_{(0)} = u^\mu = \dl^\mu{}_0$,
$\lm^\mu{}_{(1)} = \lt(0, 0, 1/{\cal R}, 0\rt)$, $\lm^\mu{}_{(2)} = \lt(0, 0, 0, 1/({\cal R} \, \sin \th)\rt)$ and
$\lm^\mu{}_{(3)} = \lt(0, 1/b, 0, 0\rt)$.
We find that for a fundamental observer at a fixed position in space, the nonzero components of the curvature tensor can
be obtained from
\be
{}^F{}R_{0101} & = & {}^F{}R_{0202} \ = \ K_1(T) \, ,
\label{RF-K1}
\nl
{}^F{}R_{0303} & = & K_2(T) \, ,
\label{RF-K2}
\nl
{}^F{}R_{3131} & = & {}^F{}R_{3232} \ = \ K_3(T) \, ,
\label{RF-K3}
\nl
{}^F{}R_{1212} & = & K_4(T) \, .
\label{RF-K4}
\ee
Thus the analogue of (\ref{R-matrix}) in this case is given by $E = {\rm diag}(K_1, K_1, K_2)$, $B = 0$, and $N = {\rm diag}(K_3, K_3, K_4)$.
Here,
\be
K_1(T) & = & -{\ddot{\cal R} \over {\cal R}} \, , \qquad \hspace{1mm} K_4(T) \ = \ {2 \over {\cal R}^2} \, \lt({1 \over 2} \, \dot{\cal R}^2 - {\cal E} \rt) \, ,
\label{K1-K4}
\nl
K_2(T) & = & -{\ddot{\cal R}' \over {\cal R}'} \, , \qquad K_3(T) \ = \ {1 \over {\cal R R}'} \, {\partial \over \partial r} \lt({1 \over 2} \, \dot{\cal R}^2 - {\cal E} \rt) \, ,
\label{K2-K3}
\ee
where $t = T$ and $r$ is simply a constant in the final expressions.
Then (\ref{F-g00})--(\ref{F-gij}) together with (\ref{Fermi-coord}) imply that the Fermi system $(T, \vec{X})$ along
the worldline of any fundamental observer has a metric of the form
\be
- \d s^2 & = &
-\lt[1 + \lt(K_1 \, \sin^2 \Th + K_2 \, \cos^2 \Th \rt)\rho^2 \rt] \d T^2 + \d \rho^2
+ \lt(1 - {1 \over 3} \, K_3 \, \rho^2 \rt) \rho^2 \, \d \Th^2
\nn
&  &{} + \lt[1 - {1 \over 3} \lt(K_4 \, \sin^2 \Th + K_3 \, \cos^2 \Th \rt)\rho^2 \rt] \rho^2 \, \sin^2 \Th \, \d \Ph^2 \, ,
\label{LTB-metric-fermi}
\ee
which is axially symmetric about the $Z$-axis (i.e. the radial $r$-direction), as expected.
Local dynamics in this spacetime is considered in Appendix~\ref{Appendix:A}.


We are particularly interested in the form of this metric in the neighbourhood of the fundamental observer at the center of
spherical symmetry $(r = 0)$.
Assuming that the spacetime manifold is smooth at $r = 0$ \cite{Bondi}, we can write
\be
{\cal R}(t,r) & = & r \, S(t) \lt[1 + {1 \over 2} \, \Dl (t) \, r + {1 \over 6} \, \Sg(t) \, r^2 + O(r^3) \rt] ,
\label{R-LTB}
\nl
{\cal E}(r) & = & {1 \over 2} \, {\cal E}''(0) \, r^2 + O(r^3) \, ,
\label{E}
\nl
M(r) & = & {1 \over 6} \, M'''(0) \, r^3 + O(r^4) \, .
\label{M}
\ee
Here, the scale factor is given by $S(t) = {\cal R}'(t,0) > 0$ and
\be
\Dl(t) & \equiv & {{\cal R}''(t,0) \over {\cal R}'(t,0)} \, , \qquad \Sg(t) \ \equiv \ {{\cal R}'''(t,0) \over {\cal R}'(t,0)} \, .
\label{Dl-Sg}
\ee
Using these approximate expressions, we find that
\be
K_1 & = & {\cal U}(T) \lt[1 + {1 \over 2} \, \eta \, r + O(r^2)\rt], \qquad K_2 \ = \ {\cal U}(T) \lt[1 + \eta \, r + O(r^2)\rt] ,
\label{K1-K2-approx}
\nl
K_3 & = & {\cal W}(T) \lt[1 + O(r)\rt], \hspace{2.1cm} K_4 \ = \ {\cal W}(T) \lt[1 + O(r)\rt] ,
\label{K3-K4-approx}
\ee
where
\be
{\cal U} & = & -{\ddot{S} \over S} \, , \qquad {\cal W} \ = \ \lt(\dot{S} \over S\rt)^2 - {{\cal E}''(0) \over S^2} \, ,
\label{U-W}
\ee
and $\eta = \lt(2 \dot{S} \dot{\Dl} + S \ddot{\Dl}\rt)/\ddot{S}$.
The Fermi metric for $r = 0$ then takes the form
\be
- \d s^2 & = & -\lt(1 + {\cal U} \, \rho^2 \rt) \d T^2 + \d \rho^2
+ \lt(1 - {1 \over 3} \, {\cal W} \, \rho^2 \rt) \rho^2 \lt(\d \Th^2 + \sin^2 \Th \, \d \Ph^2\rt).
\label{LTB-metric-fermi-approx}
\ee
It is now a simple matter to recognize that the gravitational field equations for metric (\ref{LTB-metric-fermi-approx}) result
in the analogues of equations (\ref{u+Lm1}) and (\ref{w+Lm1}) in this pressure-free case, namely,
\be
{\cal U} & = & {4 \pi G \over 3} \, \mu(t,0) - {\Lm \over 3} \, ,
\label{U+Lm1}
\nl
{\cal W} & = & {8 \pi G \over 3} \, \mu(t,0) + {\Lm \over 3} \, .
\label{W+Lm1}
\ee
Here $\mu(t,0)$ can be expressed as
\be
\mu(t,0) & = & {M'''(0) \over 8 \pi} \, {1 \over S^3(t)}
\label{mu}
\ee
by virtue of equation (\ref{dM/dr-def}).

It follows from a comparison of (\ref{U-W}) with (\ref{w-def}) that $-{\cal E}''(0)$ plays the role of spatial curvature
in these inhomogeneous models.
Moreover, $k(T) = q H^2$ in this case and it is interesting to compare this appearance of the deceleration parameter,
defined in terms of $S(t)$ in the standard manner, with the {\em effective} deceleration parameter $Q$ obtained from the
luminosity distance-redshift relation.
It has been shown \cite{Partovi} that in the general pressure-free case,
\be
d_L & = & {1 \over H} \, z + {1 \over 2H} \, (1 - Q) \, z^2 + \cdots \, .
\label{d_L1}
\ee
where the inhomogeneity parameter is given by
\be
Q - q & = & {1 \over SH^2} \, {\d \Dl(t) \over \d t} \, .
\label{Q-q}
\ee
The LTB models have recently received much attention as simple alternatives to the dark-energy models of the universe;
see, for example, \cite{Celerier2,Iguchi,Mansouri,Moffat,Apostolopoulos,Alnes} and references therein.

Equations (\ref{U-W}) and (\ref{U+Lm1})--(\ref{mu}) may be written as
\be
{1 \over 2} \, {\cal E}''(0) & = & {1 \over 2} \lt(\d S \over \d t\rt)^2 - {1 \over 6} \lt[{GM'''(0) \over S} + \Lm \, S^2 \rt],
\label{E''}
\nl
2q \, H^2 & = & H^2 - \lt[{{\cal E}''(0) \over S^2} + \Lm \rt].
\label{qH^2}
\ee
The scale factor $S(t)$ can be simply determined from (\ref{E''}) by quadratures.
In general, however, a simple analytic solution is not available.
The qualitative behaviour of $S(t)$ can be studied using the interpretation of (\ref{E''}) in terms of the radial motion
of a ``particle'' of unit mass with kinetic energy $\dot{S}^2/2$ and total energy ${\cal E}''(0)/2$.
That is, a graph of the potential energy in (\ref{E''}) versus $S$ can be used to illustrate the fact that motion
can take place only in regions where
\be
{\cal E}''(0) & > & -{1 \over 3} \lt[{GM'''(0) \over S} + \Lm \, S^2 \rt].
\label{E''-condition}
\ee
Alternatively, (\ref{E''}) can be integrated numerically with the initial condition that at the present epoch $t_0$,
$S(t_0) = R_0$, where $R_0$ is the (spatial) curvature radius.
Let us note that for ${\cal E}''(0) = 0$ and $\Lm = 0$,
\be
S & = & \lt[{3 \over 4} \, GM'''(0) \, t^2\rt]^{1/3},
\label{S-special}
\ee
so that $H = 2/(3t)$, and $q = 1/2$ follows immediately from (\ref{qH^2}); therefore, this case corresponds to the familiar
Einstein-de Sitter model.
Equation (\ref{qH^2}) is used in the next section to estimate the influence of the cosmic gravitational field on the dynamics
of the solar system.

\section{Local Dynamics}
\label{Sec5}

The expanding universe is over ten billion years old; therefore, the cosmic tidal acceleration within the solar system
is expected to be relatively very small.
The long-term cosmological evolution of an ``isolated'' gravitationally bound system in an expanding universe is beyond the scope of this work;
instead, we are interested in observable cosmological perturbations on the dynamics of the solar system.
As explained in detail in the Appendix, the influence of spatial inhomogeneities on a Keplerian binary system can be studied on
the basis of the Fermi coordinate system (\ref{LTB-metric-fermi}) associated with the LTB model.
However, the small anisotropy of the cosmic microwave background radiation implies that a useful estimate of the effect
can be obtained by ignoring any deviations from spherical symmetry.
Therefore, we consider the quasi-inertial Fermi coordinate system that can be established along the
worldline of the fundamental observer at $r = 0$.
Within this co-ordinate system associated with equation (\ref{LTB-metric-fermi-approx}), imagine the approximately elliptical
relative orbit of a Keplerian binary system that is perturbed by the cosmic tidal acceleration
\be
{\d^2 \vec{X} \over \d T^2} + {GM_0 \over \rho^3} \, \vec{X} & = & \vec{g}_{\rm cosmos} \, ,
\label{tidal-acceleration}
\ee
where at the present epoch, $\vec{g}_{\rm cosmos} = -q_0 H_0^2 \, \vec{X}$ according to equation (\ref{LTB-metric-fermi-approx}).
Here, $M_0$ is the net inertial mass of the binary system.
Let us note that if the members of the Keplerian binary were test particles, i.e. $M_0 \rightarrow 0$, then equation (\ref{tidal-acceleration})
would simply reduce to the Jacobi equation at the present epoch, since the relative speed of the particles is assumed to be negligible compared to the speed of light.
If the external tidal acceleration is turned off at any instant of time, the resulting orbit is the osculating ellipse at that
instant with eccentricity $e$ and semimajor axis $A$.
Thus, the Newtonian orbital energy of the osculating ellipse is $-GM_0/(2A)$.
On the other hand, it follows from (\ref{tidal-acceleration}) that the rate of change of this orbital energy is given by
$\vec{g}_{\rm cosmos} \cdot \vec{V}$, so that
\be
{GM_0 \over 2A^2} \, {\d A \over \d T} & = & \vec{g}_{\rm cosmos} \cdot \vec{V} \, .
\label{orbital-energy-change}
\ee

For the osculating ellipse,
\be
\vec{X} \cdot \vec{V} & = & \omK \, A^2 \, e \sqrt{1 - e^2} \, {\sin \vph \over 1 + e \, \cos \vph} \, ,
\label{X.V}
\ee
where $\omK = \lt(GM_0/A^3\rt)^{1/2}$ is the Keplerian frequency, and $\vph$ is an azimuthal angle of the orbit (``true anomaly'').
Moreover, it follows from (\ref{qH^2}) that $2q_0 = 1 - \chi_0$, where $\chi_0$ is the present value of
\be
\chi & = & {1 \over H^2} \lt[{{\cal E}''(0) \over S^2} + \Lm\rt].
\label{chi}
\ee
Putting equations (\ref{orbital-energy-change})--(\ref{chi}) together, we find that $\d A/\d T$ has the same periodicity
as the orbit and its average over an orbit vanishes; that is
\be
{\d A \over \d T} & = & - {H_0^2 A \, e \over \omK} \lt(1 - \chi_0\rt) {\sqrt{1 - e^2} \, \sin \vph \over 1 + e \, \cos \vph} \, .
\label{dA/dT}
\ee
This is in fact the Lagrange planetary equation for the semimajor axis of the osculating ellipse in this particular case.
It turns out that---on the average---the cosmological perturbations under consideration here cause a precession of the orbit in its plane,
but leave the orbit otherwise unchanged.
The Appendix should be consulted for further details about the average behaviour of the orbit in this case.

For the orbit of the Earth around the Sun,
\be
{H_0^2 A \, e \over \omK} & \approx & {1 \over 4} \times 10^{-9} \, {\rm cm/yr} \, ,
\label{H^2Ae/omega}
\ee
where $e \approx 0.02$, $A \approx 1.5 \times 10^{13}$ cm, and $H_0 \approx 70$ km s$^{-1}$ Mpc$^{-1}$.
Even if---as a result of spatial inhomogeneities or a nonzero cosmological constant---the absolute magnitude
of the deceleration parameter is enhanced by several orders of magnitude or so,
the rate of variation of the astronomical unit would still be too small to be detectable at present.
This should be contrasted with the recent reported secular increase of the astronomical unit, based on radiometric data,
amounting to about $10$ cm per year \cite{Krasinsky,Lammerzahl}.
We therefore conclude that the presence of cosmological inhomogeneities (or a cosmological constant) cannot change the
conclusion that the expansion of the universe has a negligible influence on the dynamics of the solar system.

The results of this section as well as Appendix~\ref{Appendix:A} indicate that solar-system anomalies \cite{Lammerzahl}---such as the
Pioneer anomaly---cannot be explained in terms of cosmological perturbations based upon the general
relativistic cosmological models considered in the present work.
This is particularly evident from the negligibly small magnitude of $\vec{g}_{\rm cosmos}$ over
the solar system, since for $\lt|\Xvec\rt| \sim 100$ AU, $H_0^2 \lt|\Xvec\rt| \sim 10^{-20} \ {\rm cm/s^2}$.
This is about thirteen orders of magnitude smaller than the anomalous acceleration
of Pioneer spacecraft.

\section{Discussion}
\label{Sec6}

For local systems, such as the solar system, each body is subject to the gravitational influence of the whole mass-energy content of the universe;
therefore, the relative motion of bodies is only affected by the tidal acceleration of the cosmic gravitational field.
The main purpose of this paper has been to study the general features of tidal dynamics in cosmological models.
Particular emphasis has been placed on inhomogeneous models, since spatial inhomogeneities mimic dark energy.
Indeed, we have elucidated the influence of inhomogeneities on tidal dynamics.
The results of this work could therefore be of interest in the theoretical study of the tidal interaction between galaxies.

The tidal influence of the cosmic gravitational field on the solar system can be estimated and the result turns out to be too small to be measurable
in the foreseeable future.


\begin{appendix}
\section{Local Dynamics in LTB Spacetime}
\label{Appendix:A}

Imagine a Keplerian binary system consisting of masses $m_1$ and $m_2$ with positions $\vec{X}_1$ and $\vec{X}_2$ within the
Fermi co-ordinate system associated with the LTB metric (\ref{LTB-metric-fermi}).
The ``Newtonian'' tidal potential
\be
{\cal V}_{\rm N} & = & {1 \over 2} \, K_1(T) \lt(X^2 + Y^2\rt) + {1 \over 2} \, K_2(T) \, Z^2
\label{V-N1}
\ee
is quadratic in $\vec{X}$, so that the corresponding tidal acceleration, $-\nabvec {\cal V}_{\rm N}$, is linear in $\vec{X}$.
This implies that the external cosmological perturbation on the Kepler system has the form
\be
\vec{F} & = & -K_1^0 \, \Xvec + \lt(K_1^0 - K_2^0\rt) Z \, \hat{\vec{Z}} \, ,
\label{F}
\ee
where $\Xvec = \Xvec_1 - \Xvec_2$ describes relative position.
Here, $K_1^0 = K_1(T_0)$ and $K_2^0 = K_2(T_0)$ are evaluated at the present epoch $T_0$.
It is important to emphasize that our task here is to study the observable consequences of cosmological perturbations
on local systems; therefore, the external perturbation on the binary system is evaluated at the present epoch in equation (\ref{F}).
The equation of relative motion now takes the standard form
\be
{\d^2 \Xvec \over \d T^2} + {GM_0 \, \Xvec \over \rho^3} & = & \vec{F} \, ,
\label{tidal-standard}
\ee
where $M_0 = m_1 + m_2$.
We assume that the binary system experiences small perturbations due to the tidal influence of the cosmological gravitational field.
The perturbed Kepler system (\ref{tidal-standard}) can thus be treated using standard methods of celestial mechanics \cite{Danby}.
In our approach, the study of the long-term cosmological evolution of the Keplerian binary should be based upon the solution
of (\ref{tidal-standard}) in the more general case in which $K_1(T)$ and $K_2(T)$ in $\vec{F}$ are not restricted to the
present epoch; however, such an analysis is beyond the scope of this paper.

The state of relative motion in (\ref{tidal-standard}) is given by the position and velocity at a given time $T$; on the other hand,
one could employ instead the six orbital elements of the instantaneous osculating ellipse to specify the motion.
The temporal evolution of these orbital elements are described by the Lagrange planetary equations \cite{Danby}, which are therefore
equivalent to (\ref{tidal-standard}).
For the orbital parameters, it is useful to employ Delaunay's action-angle elements $\lt(\tilde{L}, \tilde{G}, \tilde{H}, \tilde{l}, \tilde{g}, \tilde{h}\rt)$
given by
\be
\tilde{L} & = & A^{1/2} \, , \qquad \tilde{G} \ = \ \lt[G M_0 A \lt(1 - e^2\rt)\rt]^{1/2} \, , \qquad \tilde{H} \ = \ \tilde{G} \cos i \, ,
\label{LGH}
\nl
\tilde{l} & = & \vth - e \, \sin \vth \, , \qquad \tilde{g} \ = \ {\rm argument \ of \ the \ pericentre} \, ,
\qquad \tilde{h} \ = \ {\rm longitude \ of \ the \ ascending \ node}.
\label{l}
\ee
%
Here, $A$ is the semimajor axis of the osculating ellipse, $e$ is its eccentricity, $i$ is the orbital inclination, $\vth$
is the eccentric anomaly, and $\tilde{l}$ is the mean anomaly.
Moreover, $\tilde{G}$ is the magnitude of the orbital angular momentum vector and $\tilde{H}$ is its $Z$-component.
Along the osculating ellipse, the radial position $\rho$ can be expressed in terms of the true anomaly $\vph$ and eccentric anomaly $\vth$,
respectively, as follows
\be
\rho & = & {A \lt(1 - e^2\rt) \over 1 + e \, \cos \vph} \, , \qquad \rho \ = \ A \lt(1 - e \, \cos \vth \rt).
\label{rho}
\ee
In equation (\ref{LGH}) and throughout, we consider only positive square roots.
To express the dynamical equations in terms of Delaunay's elements, it is necessary to decompose the perturbing acceleration
$\vec{F}$ in terms of an orthonormal frame field adapted to the osculating ellipse.
That is, we can write
\be
\vec{F} & = & F_{\rho} \, \hat{\vec{\rho}} + F_s \, \hat{\vec{s}} + F_n \, \hat{\vec{n}} \, ,
\label{F1}
\ee
in terms of its radial, sideways, and normal components.
Here $\hat{\vec{\rho}} = \vec{X}/\rho$ is the radial unit vector, $\hat{\vec{s}} = \hat{\vec{n}} \times \hat{\vec{\rho}}$, and
$\hat{\vec{n}}$ is the unit vector in the direction of the orbital angular momentum $\tilde{\vec{G}}$;
that is, $\tilde{\vec{G}} = \tilde{G} \, \hat{\vec{n}}$.
Thus, $\hat{\vec{\rho}}$ and $\hat{\vec{s}}$ are in the instantaneous orbital plane of the osculating ellipse,
while $\hat{\vec{n}}$ is normal to it.

The equations of motion in terms of Delaunay's elements are then given by
\be
{\d \tilde{L} \over \d T} & = & {\tilde{L}^3 \over \tilde{G}} \lt[F_{\rho} \, e \, \sin \vph + F_s \lt(1 + e \, \cos \vph\rt) \rt],
\label{dL/dT}
\nl
{\d \tilde{G} \over \d T} & = & \rho \, F_s \, ,
\label{dG/dT}
\nl
{\d \tilde{H} \over \d T} & = & \rho \lt[F_s \, \cos i - F_n \, \sin i \, \cos \lt(\vph + \tilde{g}\rt) \rt],
\label{dH/dT}
\nl
{\d \tilde{l} \over \d T} & = & \omega_{\rm K} + {\rho \over \omega_{\rm K} A^2 e} \lt[F_{\rho} \lt(-2e + \cos \vph + e \cos^2 \vph\rt)
- F_s \lt(2 + e \cos \vph\rt)\sin \vph \rt],
\label{dl/dT}
\nl
{\d \tilde{g} \over \d T} & = & - {\rho F_n \over \tilde{G}} \, {\cos i \over \sin i} \, \sin \lt(\vph + \tilde{g}\rt)
+ {\lt(1 - e^2\rt)^{1/2} \over \omega_{\rm K} A e} \lt(-F_{\rho} \, \cos \vph + F_s \, {2 + e \cos \vph \over 1 + e \cos \vph} \, \sin \vph \rt),
\label{dg/dT}
\nl
{\d \tilde{h} \over \d T} & = & {\rho F_n \over \tilde{G}} \, {\sin \lt(\vph + \tilde{g}\rt) \over \sin i} \, ,
\label{dh/dT}
\ee
where $\omega_{\rm K} = \lt(GM_0\rt)^{1/2}/\tilde{L}^3$.
A direct derivation of these equations is essentially contained in Appendix~B of \cite{Chicone4}.

To find the explicit form of $F_{\rho}$, $F_s$, and $F_n$ in terms of Delaunay's elements, we note that \cite{Chicone4}
\be
X & = & \rho \lt[\cos \tilde{h} \, \cos \lt(\vph + \tilde{g}\rt) - \sin \tilde{h} \, \cos i \, \sin \lt(\vph + \tilde{g}\rt) \rt],
\label{X}
\nl
Y & = & \rho \lt[\sin \tilde{h} \, \cos \lt(\vph + \tilde{g}\rt) + \cos \tilde{h} \, \cos i \, \sin \lt(\vph + \tilde{g}\rt) \rt],
\label{Y}
\nl
Z & = & \rho \sin i \, \sin \lt(\vph + \tilde{g}\rt).
\label{Z}
\ee
Moreover \cite{Chicone4},
\be
\hat{\vec{n}} & = & \lt(\sin \tilde{h} \, \sin i, - \cos \tilde{h} \, \sin i, \cos i \rt),
\label{n-hat}
\nl
\hat{\vec{s}} & = & \lt(-\cos \tilde{h} \, \sin \lt(\vph + \tilde{g}\rt) - \sin \tilde{h} \, \cos i \, \cos \lt(\vph + \tilde{g}\rt), \rt.
\nn
& &{} - \lt. \sin \tilde{h} \, \sin \lt(\vph + \tilde{g}\rt) + \cos \tilde{h} \, \cos i \, \cos \lt(\vph + \tilde{g}\rt), \sin i \, \cos \lt(\vph + \tilde{g}\rt) \rt).
\label{s-hat}
\ee
A straightforward calculation using (\ref{F}) reveals that
\be
F_\rho & = & - K_1^0 \, \rho + \lt(K_1^0 - K_2^0\rt) \rho \sin^2 i \, \sin^2 \lt(\vph + \tilde{g}\rt),
\label{F-rho}
\nl
F_s & = & \lt(K_1^0 - K_2^0\rt) \rho \sin^2 i \, \sin \lt(\vph + \tilde{g}\rt) \, \cos \lt(\vph + \tilde{g}\rt),
\label{F-s}
\nl
F_n & = & \lt(K_1^0 - K_2^0\rt) \rho \sin i \, \cos i \, \sin \lt(\vph + \tilde{g}\rt).
\label{F-n}
\ee
Substituting these results in equations (\ref{dL/dT})--(\ref{dh/dT}) and averaging over the ``fast'' orbital motion, one can
determine the ``slow'' evolution of the orbit under cosmological perturbations in this case.

The weak external perturbation (\ref{F}) naturally splits into two terms:  one that is proportional to $K_1^0$
and the other proportional to $K_1^0 - K_2^0$.
The influence of these on the orbit can be analysed separately and the results can be superimposed in accordance with our linear
perturbation scheme.
Let us therefore first consider a tidal perturbation of the form $-K_1^0 \, \Xvec$.
It turns out that this problem has already been solved in a different context \cite{Kerr}; in fact, the details of the averaging
procedure are given in section~IV of \cite{Kerr} for a perturbing acceleration of the form $\lm \Xvec$, where $\lm$ is a constant.
It is interesting to note that in \cite{Kerr}, $\lm = \Lm c^2/3$, where $\Lm$ is the cosmological constant associated with the
Kerr-de Sitter spacetime.
It is shown in \cite{Kerr} that the orbit is planar and its semimajor axis and eccentricity do not change on the average;
however, the pericentre precesses with frequency $3 \lm \lt(1 - e^2\rt)^{1/2} \hat{\vec{n}}/\lt(2 \omega_{\rm K}\rt)$.
Thus the net average effect of the first term in (\ref{F}) on the orbit is to generate a pericentre precession of frequency
$-3 K_1^0 \lt(1 - e^2\rt)^{1/2} \hat{\vec{n}}/\lt(2 \omega_{\rm K}\rt)$.

The second term in (\ref{F}) is a perturbing acceleration in the $Z$-direction, which corresponds to the {\em radial} direction
in the standard form of the LTB spacetime.
Inspection of equations (\ref{dL/dT})--(\ref{dh/dT}) reveals that $\d \tilde{H}/ \d T = 0$, hence the $Z$-component of the orbital
angular momentum remains unchanged; moreover, there is clearly no effect to linear order if the orbit lies in the $(X, Y)$ plane.
Hence we assume $i \neq 0$.
It is then straightforward to average the right-hand sides of the remaining equations over the ``fast'' motion with frequency
$\omega_{\rm K} = 2 \pi / T_{\rm K}$ such that for a function $\cal F$, $\lt\langle {\cal F} \rt\rangle = T_{\rm K}^{-1} \int_0^{T_{\rm K}} {\cal F} \, \d t$,
or equivalently,
\be
\lt\langle {\cal F} \rt\rangle & = & {\lt(1 - e^2\rt)^{3/2} \over 2 \pi} \, \int_0^{2\pi} {{\cal F} \, \d \vph \over \lt(1 + e \, \cos \vph\rt)^2} \, ,
\label{<F>}
\ee
using the unperturbed orbit.
The resulting integrals can be evaluated in principle, but it is simpler to express the main results for a slightly eccentric orbit.
We find that
\be
\lt\langle {\d A \over \d T} \rt\rangle & = & O\lt(e^2\rt) \, , \qquad \lt\langle {\d e \over \d T} \rt\rangle \ = \ O\lt(e\rt) ,
\label{<dA/dT>,<de/dT>}
\nl
\lt\langle {\d \tilde{g} \over \d T} \rt\rangle & = & - {\lt(K_1^0 - K_2^0\rt) \over 2 \omega_{\rm K}}
\lt[\cos^2 i + \lt(\cos 2 \tilde{g} - 3 \, \sin^2 \tilde{g}\rt) \sin^2 i \rt] + O\lt(e\rt) ,
\label{<dg/dT>}
\nl
\lt\langle {\d \tilde{h} \over \d T} \rt\rangle & = & {\lt(K_1^0 - K_2^0\rt) \cos i \over 2 \omega_{\rm K}} + O\lt(e^2\rt) .
\label{<dh/dT>}
\ee
Thus the main effects here are the precessions of the pericentre and the ascending node with frequencies given by (\ref{<dg/dT>})--(\ref{<dh/dT>}).
The net average effect of (\ref{F}) on the orbit is mainly the precession of the orbit in its own plane as well as about the $Z$-axis
(i.e. the LTB radial direction); the latter motion occurs with frequency $\lt(K_1^0 - K_2^0\rt) \cos i/\lt(2 \omega_{\rm K}\rt)$.

Finally, let us restrict our general off-centre LTB treatment to a situation where the fundamental observer is very close
to the centre of spherical symmetry.
Indeed, the maximum anisotropy in the temperature of the cosmic background radiation is at $0.002$ level due to our motion
relative to this radiation bath; therefore, within the LTB model a reasonable estimate may be obtained by assuming $r = 0$
as in section~\ref{Sec5}.
Indeed, equation~(\ref{K1-K2-approx}) implies that $K_1 - K_2 \rightarrow 0$ for $r \rightarrow 0$.
It follows that the cosmological perturbations on average leave the shape of the orbit and the orientation of the orbital plane
unchanged, but cause a pericentre precession of frequency $-3 q_0 \, H_0^2 \lt(1 - e^2\rt)^{1/2} \hat{\vec{n}}/\lt(2 \omega_{\rm K}\rt)$.
For the motion of the Earth around the Sun, $H_0^2/\omega_{\rm K} \approx 10^{-21}/$yr, which is some fourteen orders of magnitude
smaller than the Einstein pericentre precession.
Thus the cosmological pericentre precession does not appear to be detectable in the foreseeable future.
This conclusion applies to all of the cosmological models considered in this paper, so long as any deviation
from isotropy about the Keplerian system can be neglected.

\end{appendix}

\end{document}